# Investigations of a Two-Phase Fluid Model

B. T. Nadiga

Theoretical Division, Los Alamos National Lab.,

Los Alamos, NM 87545

S. Zaleski

Laboratoire de Modélisation en Mécanique, URA 229 CNRS

Univ. P. et M. Curie (Paris 6), 4 place Jussieu

75252 Paris Cedex 05, France

**Abstract**

We study an interface-capturing two-phase fluid model in which the interfacial tension is modelled as a volumetric stress. Since these stresses are obtainable from a Van der Waals-Cahn-Hilliard free energy, the model is, to a certain degree, thermodynamically realistic. Thermal fluctuations are not considered presently for reasons of simplicity. The utility of the model lies in its momentum-conservative representation of surface tension and the simplicity of its numerical implementation resulting from the volumetric modelling of the interfacial dynamics. After validation of the model in two spatial dimensions, two prototypical applications—instability of an initially high-Reynolds-number liquid jet in the gaseous phase and spinodal decomposition in a liquid-gas system— are presented.

1. Introduction

The complexity of the physics involved at interfaces precludes any simple but comprehensive mathematical model of it and thus the proper simulation of interfaces is a difficult problem both from the physical and the computational points of view. Nonetheless, in several circumstances, it is useful to have a model of interfaces that allows for a degree of microscopic realism without being constrained by the full details that arise when interface physics are modelled in a way consistent with all the basic principles of thermodynamics and particle mechanics. Thus the utility of models that capture only some important features of the real world, but are computationally easy to deal with.

Our model is built by adding to the Navier-Stokes equations a stress tensor $\sigma^{(1)}$ that is derived from the van der Waals-Cahn-Hilliard free energy and results in spontaneous phase separation and surface tension. There is a single species of particles, of density $\rho$, with the usual continuity equation. The van der Waals equation of state ensures the presence of two basic states, a "liquid state" of density $\rho_L$ and a "gas" state of density $\rho_G$. For simplicity, the temperature in the equation of state is fixed. While this assumption is physically justifiable in specific flow situations, it simplifies the description of the flow by immediately decoupling the energy equation from the mass and momentum equations.

An important feature of our model is that there are special steady state solutions of the model mass and momentum conservation equations that connect a gas phase to a liquid phase through a smoothly varying density profile of thickness $\xi$ . Such an interface is an equilibrium state of the Cahn-Hilliard model and this ensures that we describe scales smaller than the physical thickness of the interfaces with some degree of realism. Because our model tends to create and maintain such a phase separated state, two fundamental physical phenomena are captured: (i) When initial conditions involve only thin interfaces (*i.e.*, the hydrodynamic scales, radii of curvature etc... are much larger than $\xi$), and densities are either $\rho_L$ or $\rho_G$ away from interfaces, the model simulates a gas-liquid flow with a free interface between the two phases. In this regime the flow approaches real incompressible two-phase flow when the Mach number is small. (ii) When initial conditions impose densities in the unstable region the phases will spontaneously separate. (It should be noted that in the early stages of spinodal decomposition the phase separating flow is far from being incompressible.)

Our motivation to study this kind of a model lies in (i) the existence of a spinodal decomposition regime, (ii) the fact that the model lends itself to a simple and robust numerical implementation, and (iii) that it is a stepping stone to the more complete thermodynamic model summarized in the Appendix. Interestingly, the first two characteristics of the model are shared in common with the immiscible lattice-Boltzmann models [Appert et al. 1995; Alexander, Chen and Grunau, 1993], which are themselves not thermodynamically correct. We note that, unlike most lattice-Boltzmann models (with the exception of the recent model by Swift, Osborn, and Yeomans, (1995)), the stresses inside an equilibrium interface (thus at uniform temperature) are correct in our model. Further, in lattice-Boltzmann models, the tangential velocities are not continuous when $\rho_L \neq \rho_G$, [see Rothman and Zaleski, 1994, Ginzburg 1994]. Thus, we may say that our model, while preserving the advantages of the lattice-Boltzmann models, eliminates some of its drawbacks.

Another analogy may be drawn with some surface tension schemes used in other interface simulation methods. As in [Brackbill et al. 1993] surface tension is made easier to represent numerically because the surface tension stresses are spread continuously over an interface region of finite thickness. Moreover, as in [Lafaurie et al. 1994] the surface tension terms conserve momentum exactly.

In section 2, we present the model, interpret the momentum flux components specific to the interface and obtain the surface tension coefficient. In section 3 we describe briefly the numerical method. In section 4, we present some numerical simulations of the model: we show first that the equilibrium densities and pressures on the liquid-gas interface are predicted by the model in accordance with the Maxwell equal area rule. We also show that Laplace's law is recovered. Next we present the simulation of the instability of a high Reynolds number liquid jet in the gaseous phase at two different values of the surface tension. Finally an example simulation of isothermal spinodal decomposition with the model is presented.

## 2. The model and the Korteweg interfacial stresses

We start with the mass and momentum conservation equations:

$$\partial_t \rho + \partial_j(\rho u_j) = 0, \tag{1}$$

$$\partial_t(\rho u_i) + \partial_j(\rho u_i u_j) = \partial_j \sigma_{ij}. \tag{2}$$

We decompose the stresses $\sigma_{ij}$ as $\sigma_{ij} = -p_0 \delta_{ij} + \sigma_{ij}^{(v)} + \sigma_{ij}^{(1)}$, where $p_0$ is the bulk pressure assumed to depend only on the density $\rho$ through an equation of state $p_0 = p_0(\rho)$, and $\sigma_{ij}^{(v)}$ is the viscous stress tensor

$$\sigma_{ij}^{(v)} = \mu \left( \partial_j u_i + \partial_i u_j - \frac{2}{3} \nabla \cdot \mathbf{u} \delta_{ij} \right). \tag{3}$$

The stresses $\sigma_{ij}^{(1)}$ depend on the density gradient and is the key feature of our model. These stresses may either be postulated in the manner of Korteweg (1901) or derived thermodynamically from the Van der Waals-Cahn-Hilliard free energy (see references in the Appendix as well as the historical review of D. D. Joseph (1990))

$$\sigma_{ij}^{(1)} = \lambda \left[ \left( \frac{1}{2} |\nabla \rho|^2 + \rho \nabla^2 \rho \right) \delta_{ij} - \partial_i \rho \partial_j \rho \right]. \tag{4}$$

where $\lambda$ is a constant parameter that controls the strength of the surface tension effect. While several equations of state $p_0(\rho)$ will serve our purpose, for definiteness, and in the numerical simulation that follows, we choose a van der Waals equation

$$p_0(\rho') = p_c \rho' \theta' \left( \frac{8}{3 - \rho'} - \frac{3\rho'}{\theta'} \right) \tag{5}$$

where

$$p_c = \frac{3}{8} \rho_c R_g \theta_c, \qquad \rho' = \frac{\rho}{\rho_c}, \qquad \theta' = \frac{\theta}{\theta_c}$$

and $R_g = R/M$ is the universal gas constant divided by the molecular weight and the subscript $c$ refers to the critical values (see also the Appendix, Eq. (15)). This equation of state allows for the existence of two phases of different densities: there is a range of densities over which the modeled fluid is mechanically unstable, i.e., an increase in density results in a decrease in pressure, leading to a separation into two different phases. Unlike gas-dynamical models in which the mass and momentum balance equations must be supplemented with an equation for either energy or entropy, an "artificially compressible" model such as ours needs no further equation. Our model is thus entirely described by Eqs. (1)-(5).

– 5–Considering the stress tensor

$$\sigma_{ij}^{(1)} = -\lambda \left(\rho \nabla^2 \rho \delta_{ij} + T_{ij}\right), \tag{6}$$

where

$$T = \begin{pmatrix} (\partial_x^2 \rho - \partial_y^2 \rho)/2 & \partial_x \rho \partial_y \rho \\ \partial_x \rho \partial_y \rho & -(\partial_x^2 \rho - \partial_y^2 \rho)/2 \end{pmatrix}.$$

The first term in Eq. (6) can be thought of as a pressure term acting to smooth gradients in the density field, with no contributions to the surface tension. The second term $T_{ij}$ is traceless and acts to simulate the surface tension: consider a planar interface along the $x$-direction. The only nonzero gradients are along the $y$-direction and we find that

$$T = \begin{pmatrix} -\partial_y^2 \rho/2 & 0 \\ 0 & \partial_y^2 \rho/2 \end{pmatrix}.$$

The mechanical force per unit length along the $y$-direction is $\lambda \int_{-\infty}^{\infty} (T_{yy} - T_{xx})\,dy$. Identifying it with the surface tension coefficient $\alpha$ gives us

$$\alpha = \lambda \int_{-\infty}^{\infty} \left(\frac{d\rho}{dy}\right)^2 dy, \tag{7}$$

where $\alpha$ is the surface force per unit length.

## 3. The Numerical Scheme

The numerical scheme consists of a two step, MacCormack–type predictor–corrector methodology [Peyret and Taylor, 1983] for each of the equations. For convenience, the 2D mass, momentum, and energy conservation equations can be written in the form

$$\partial_t \mathbf{f} + \partial_1 \left(\mathbf{F}[\mathbf{f}]\right) + \partial_2 \left(\mathbf{G}[\mathbf{f}]\right) = 0, \tag{8}$$

where the square parentheses indicate functional dependence, specifically

$$F[\mathbf{f}] = F\left(\mathbf{f}, \partial_1 \mathbf{f}, \partial_2 \mathbf{f}, \nabla \rho\right).$$

At any grid point the predictor step

$$\begin{aligned}
\widehat{\mathbf{f}} = \mathbf{f}^n &- \Delta t \frac{\Delta_{bck}}{\Delta x_1} \left[\mathbf{F}\left(\mathbf{f}^n, \frac{\Delta_{fwd}\mathbf{f}^n}{\Delta x_1}, \frac{\Delta_{ctr}\mathbf{f}^n}{\Delta x_2}, \nabla^2_{ctr}\rho^n\right)\right] \\
&- \Delta t \frac{\Delta_{bck}}{\Delta x_2} \left[\mathbf{G}\left(\mathbf{f}^n, \frac{\Delta_{fwd}\mathbf{f}^n}{\Delta x_2}, \frac{\Delta_{ctr}\mathbf{f}^n}{\Delta x_1}, \nabla^2_{ctr}\rho^n\right)\right],
\end{aligned} \tag{9}$$

is followed by the corrector step

$$\mathbf{f}^{n+1} = \frac{1}{2}(\mathbf{f}^n + \widehat{\mathbf{f}}) - \frac{1}{2}\Delta t \frac{\Delta_{fwd}}{\Delta x_1} \left[ \mathbf{F} \left( \widehat{\mathbf{f}}, \frac{\Delta_{bck}\widehat{\mathbf{f}}}{\Delta x_1}, \frac{\Delta_{ctr}\widehat{\mathbf{f}}}{\Delta x_2}, \nabla^2_{ctr}\widehat{\rho} \right) \right] \\ - \frac{1}{2}\Delta t \frac{\Delta_{fwd}}{\Delta x_2} \left[ \mathbf{G} \left( \widehat{\mathbf{f}}, \frac{\Delta_{bck}\widehat{\mathbf{f}}}{\Delta x_2}, \frac{\Delta_{ctr}\widehat{\mathbf{f}}}{\Delta x_1}, \nabla^2_{ctr}\widehat{\rho} \right) \right], \qquad (10)$$

where $\Delta_{fwd}$ stands for the forward difference, $\Delta_{bck}$ for the backward difference in the relevant direction, and $\nabla^2_{ctr}$ for the center-differenced Laplacian operator.

## 4. Numerical Simulations

There are three dimensionless parameters of interest in the problem: the Reynolds number, the Weber number, and a Mach number:

$$Re = \frac{\rho_c V L}{\mu}, \qquad We = \frac{V^2 L^2}{\rho_c \lambda}, \qquad Ma_c = \frac{V}{\sqrt{R_g T_c}}. \qquad (11)$$

The number $Ma_c$ appears immediately in the non-dimensional form of the pressure, (5). However, it is physically more appealing to consider the numbers $Ma(\rho) = V/c_s$ where $c_s^2 = dp_0/d\rho$ is the sound speed squared, and depends on the density $\rho$.

Towards validating the numerical model, we first present the results of a few test cases. In Fig. 1, at a given temperature, the solid line gives the densities of the two coexisting phases calculated using the Maxwell equal area construction. The symbols are the densities of the liquid and gas phases obtained from the numerical simulations. The simulation consisted of equilibration of a flat interface between the high and low density phases on a 64 × 64 grid. The Reynolds number in (11) (based on the grid spacing $\Delta x$) was set at 2.0 and the Weber number (also based on $\Delta x$) was set at 1.0. Using formula (7), the value of the nondimensional surface tension coefficient $(\alpha_{dim}\Delta x/(\rho_c^2\lambda)$, where $\alpha_{dim}$ is the dimensional surface tension coefficient) is calculated for future reference to be $\alpha(\theta/\theta_c = 0.85, We = 1.0) = 0.47$ and $\alpha(\theta/\theta_c = 0.85, We = 0.6) = 0.64$.

Next we verify Laplace's law for surface tension. The simulations here consisted of letting go to equilibrium drops of the liquid phase suspended in the gaseous phase. The computational domain in this case was either 256×256 or 64×64 and the Reynolds number was set at 2.0 (again based on $\Delta x$).

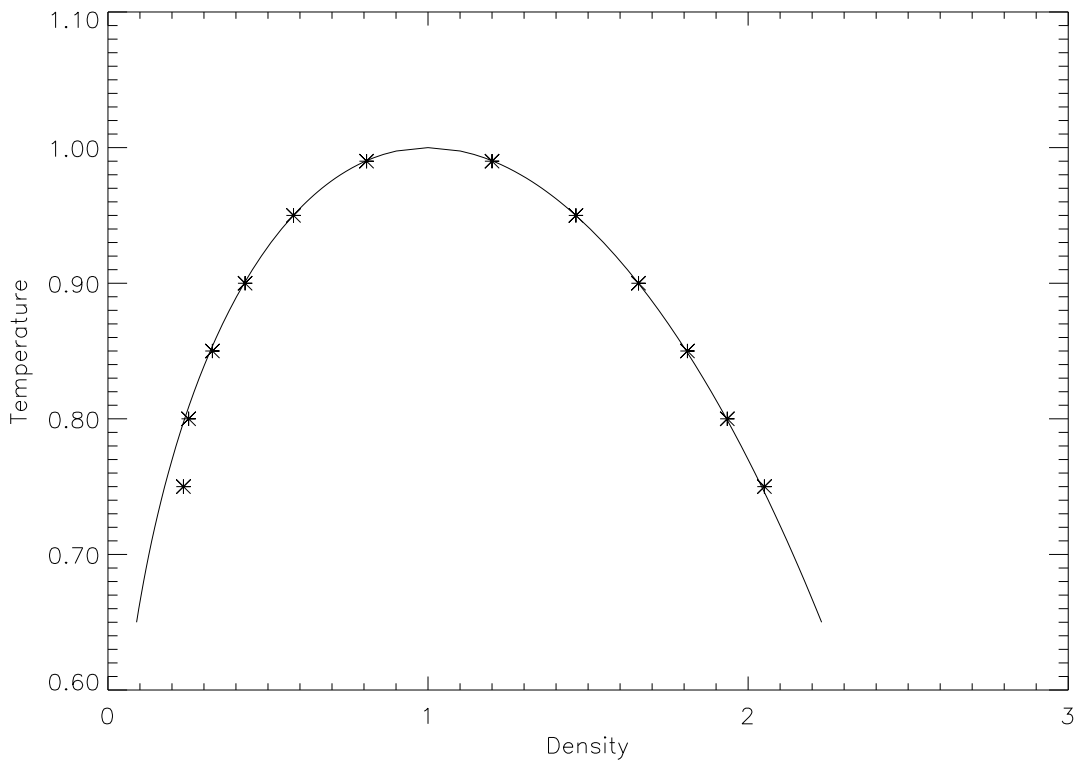

Fig. 1 The densities of the coexisting phases at different temperatures. The solid line results from a thermodynamic calculation in which the Gibbs free energies of the two phases at the given temperature are equated. The symbols are results from the numerical simulations.

The Weber number was either 0.6 or 1.0 and the surface tension coefficient $\alpha$ calculated from the flat interface simulations were used to plot $\alpha/R$ on the $x$-axis and the measured pressure difference between the bulk liquid and gas phases $\Delta p$ is plotted on the $y$-axis in Fig. 2. The deviation from the least squares fit is within 5%. The linearity of the plot verifies Laplace's law and the consistency in the model of the surface tension, noting that the surface tension coefficient was obtained from the flat interface simulations. We next present two simple applications of the model—a case of instability of a high Reynolds number liquid jet in the gaseous phase and an instance of isothermal spinodal decomposition.

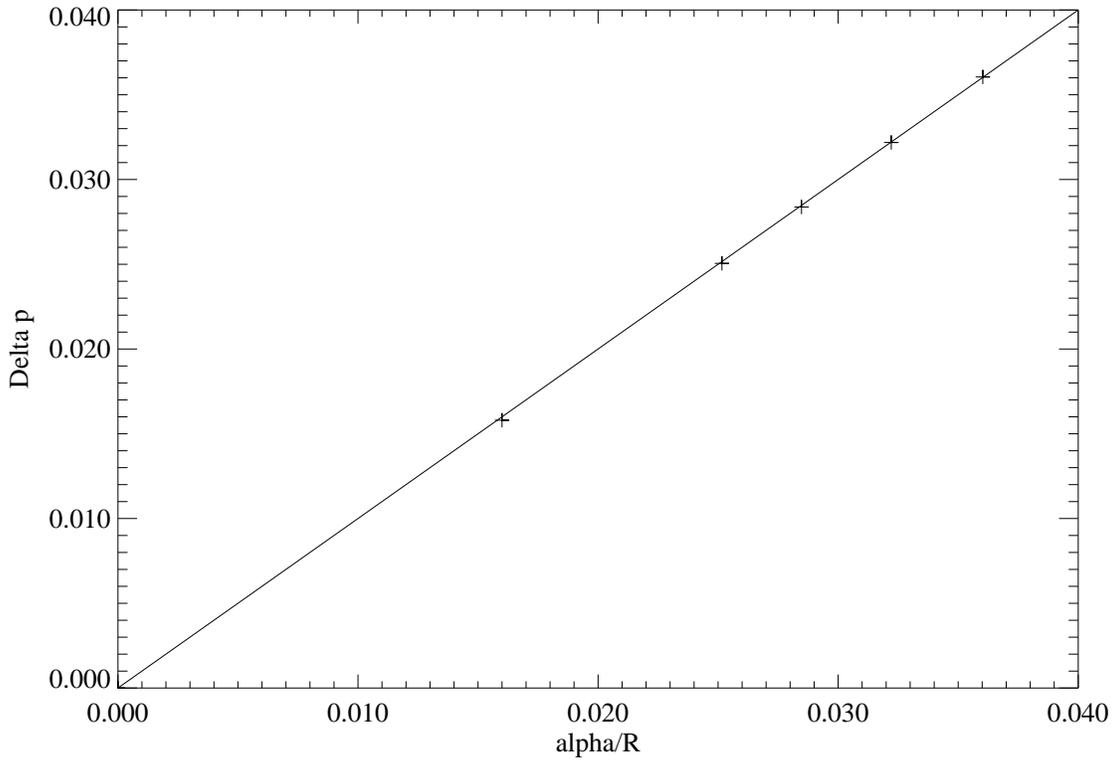

FIG. 2 Verification of Laplace's law. On the $x$-axis we plot $\frac{\alpha}{R}$, where $R$ is the radius of the drop and $\alpha$ is the value of the surface tension coefficient calculated from the above simulation of flat interfaces. On the $y$-axis is the numerically measured pressure difference between the inside and the outside of the drop. The solid line is the least-squares linear fit and has a slope of 1.037 with a standard deviation of 0.0008.

### 4.1 Instability of a planar jet

We consider the evolution of a high Reynolds number planar jet of the liquid phase in the gaseous phase. We note that there are very few simulations of two-phase flows at high Reynolds number considering their computational difficulty. Fig. 3 shows the shape of the jet at times 0, 200, 400, 800, and 1000 nondimensional units. Initially (t=0), the Reynolds number of the jet based on the diameter is about 800, the Weber number, also based on the diameter is about 80, and $Ma_c$ is about 0.23. 512 grid points are used along the axis of the jet and 128 grid points transverse to it. The boundary conditions are doubly periodic and the computations are isothermal. The instability of the jet is seen as the unsteady deformation of its shape. Note that in the absence of forcing, the energy of the jet is constantly decaying with time. In a second experiment, the same initial jet configuration is used, but the surface tension is increased by decreasing the Weber number to about 40. The snapshots of the jet at the same times as in Fig. 3 are shown for the case with the increased surface tension in Fig. 4. The simulation clearly show how the increased surface tension tends to stabilize the jet.

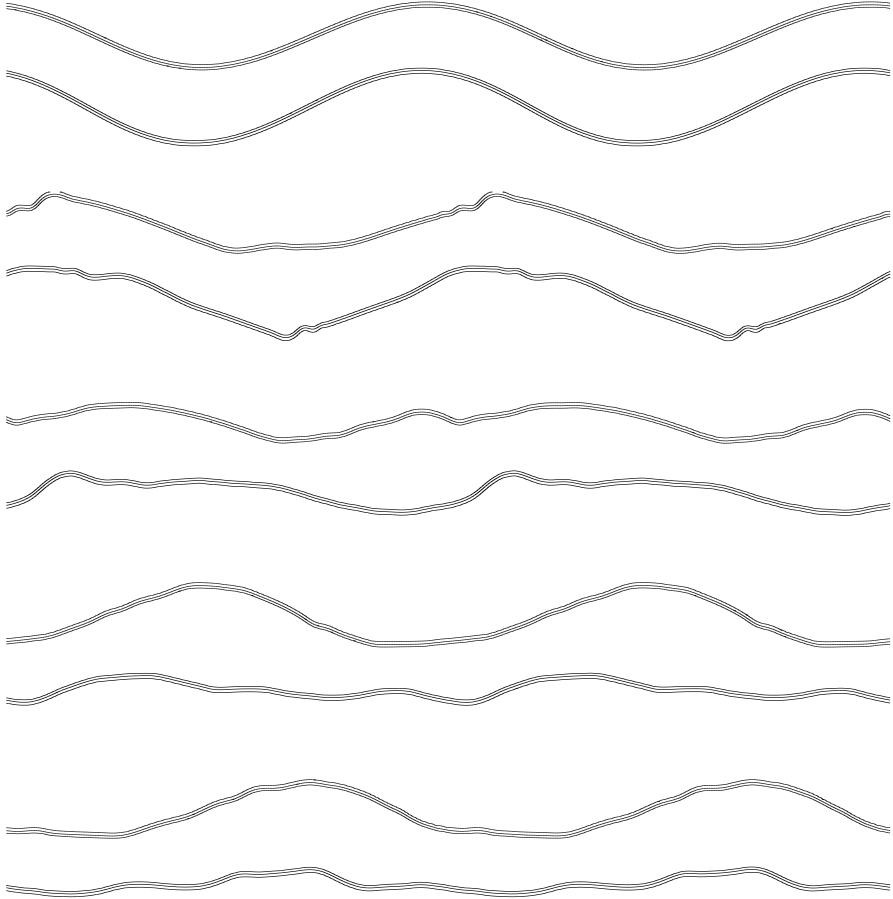

FIG. 3 Instability of a planar liquid jet in the gas phase at an initial Reynolds number (based on the diameter) of about 800 and a Weber number (based on the diameter) of about 80. The snapshots of the liquid jet at time 0, 200, 400, 800, and 1000 are shown from top to bottom for this isothermal simulation.

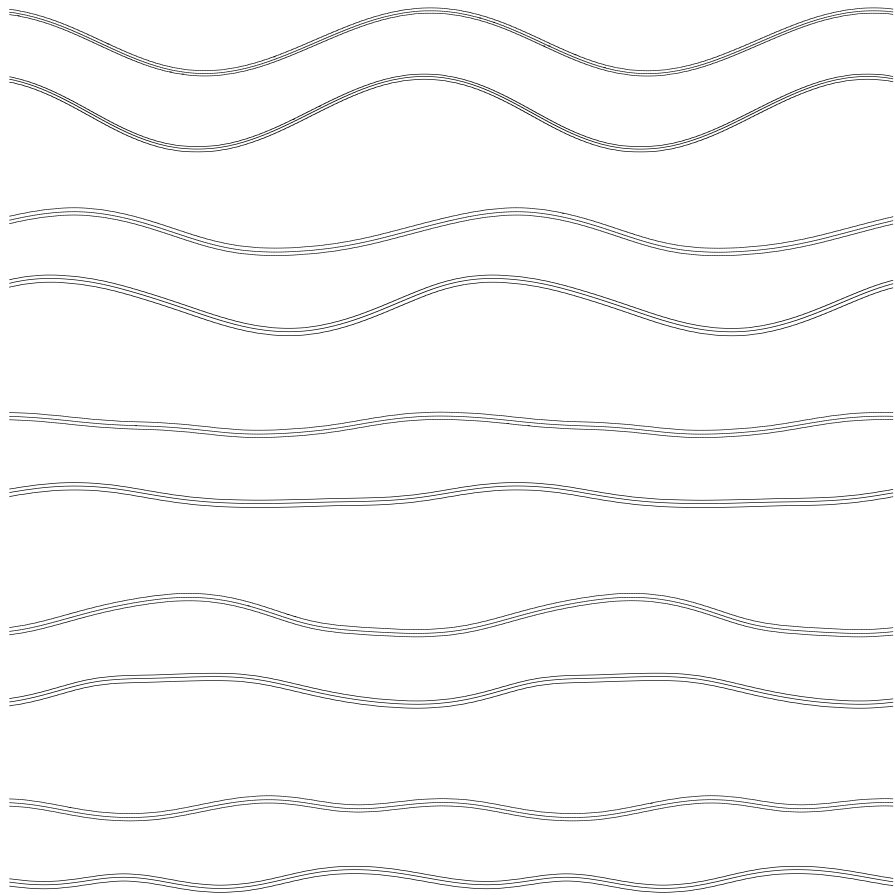

FIG. 4 The same setup as in Fig. 3, but the surface tension is increased by a factor of 2 from that case. The snapshots are at the same times as before, and the increased stability of the jet is clear.

## 4.2 Spinodal Decomposition

When a solid or fluid is rapidly quenched from above the critical point where there is only one homogeneous phase to below the critical point where two or more different phases may coexist (within the spinodal), the homogenous phase is unstable and therefore separates spontaneously into domains occupied by the coexisting phases. After that, the average size of the domains tends to increase in an effort to decrease the interfacial energy. Such spontaneous formation of domains and their subsequent coarsening is called spinodal decomposition. The rate of growth of the average size of these domains is both of theoretical interest and of practical value. When the system undergoing spinodal decomposition is a fluid system, the advection of fluid particles leads to an enhanced growth rate as compared to — say — alloys where there are no advective processes and the growth is limited by diffusion. While the precise mechanisms and the growth exponents are not well understood in the presence of fluid dynamics, particularly when aspects like compressibility are concerned, it is expected that there exist a few universality classes characterizing the growth of domains in such systems. This expectation is based in part on the success of scaling theories in conjunction with numerical and experimental studies in binary alloys [Gunton, San Miguel, and Sahni, 1983; Rogers, Elder, and Desai, 1988; and references therein] and some such similar successes in immiscible binary fluid systems [Ma, Maritan, Banavar, & Koplik, 1992; Farrell and Valls, 1990; Furukawa, 1985; and references therein]. The numerical study of spinodal decomposition in liquid-vapor systems however has not received much attention (in contrast to binary alloy and immiscible binary fluid systems) mainly because of a lack of models including all of the fluid dynamics and the energetics (exceptions are Langevin fluid models [Farrell and Valls, 1990]). The present compressible model is simple and complete except for the effects of temperature variations and the release of latent heat.

The initial condition is a uniform state at a supercritical temperature with a noise of amplitude $0.2\rho_c$ in the density field superposed over an average density of $1.06344\rho_c$. The system is then quenched to $0.85\theta_c$. After this temperature quench, the homogenous phase finds itself in a region of mechanical instability (pressure decreases with increasing density) and therefore separates. Fig. 5 shows the snapshots at times 25, 50, 250 and 500 nondimensional time units of the spinodal decomposition following the temperature quench. A doubly periodic $512 \times 512$ domain was used. At the final time shown (t=500) in Fig.5, $Ma_c$ is 1.52, based on the velocity averaged over the domain. The Reynolds number in the dense phase, based on the average domain size, is about 95, and the Weber number (again based on the average domain size) is about 120.

By considering the circularly-averaged two-point correlation function (the second-order structure function) of the density at a given time, the average domain size is estimated as the first zero crossing of the correlation function. The domain growth with time is shown as symbols on a log-log plot in Fig. 6. Here the data is obtained by an ensemble average over 12 different runs in which only the seed of the random number generator was varied to initialize the noise in the density field. For the domain sizes lying between $20\Delta x$ and $128\Delta x$ (the computational domain was $512\Delta x \times 512\Delta x$), a least-squares linear fit gave us a slope of 0.70 with a standard deviation of 0.01. The data for which the power law growth was estimated (the late stage) is plotted as '+' and the rest of the data as open squares. The region of Fig.6 over which the data was fit with a power law is replotted to show the fit more clearly. We plan to present a more complete study of this problem elsewhere, but we point out that in this two-dimensional isothermal simulation where inertial effects are important (late stage) the 0.70 growth exponent is close to the higher values of the exponents that have been observed [Farrell and Valls, 1990].

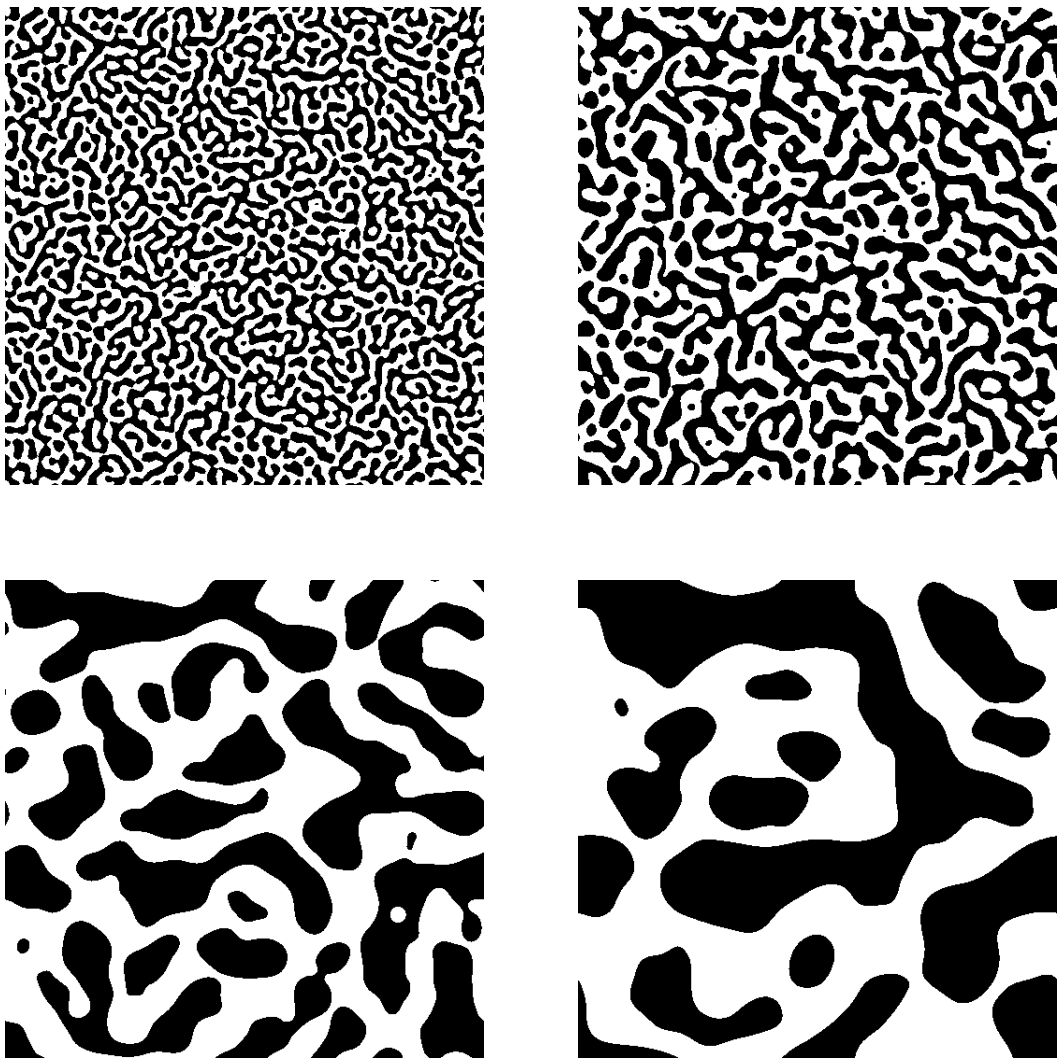

FIG. 5 Spinodal decomposition in the van der Waals gas following a temperature quench at time 0. The four snapshots of the density field are at times 25, 50, 250, and 500 nondimensional time units. The average density is 1.063 and the density of the coexisting liquid and gas phases are respectively 0.319 and 1.804 (across a flat interface).

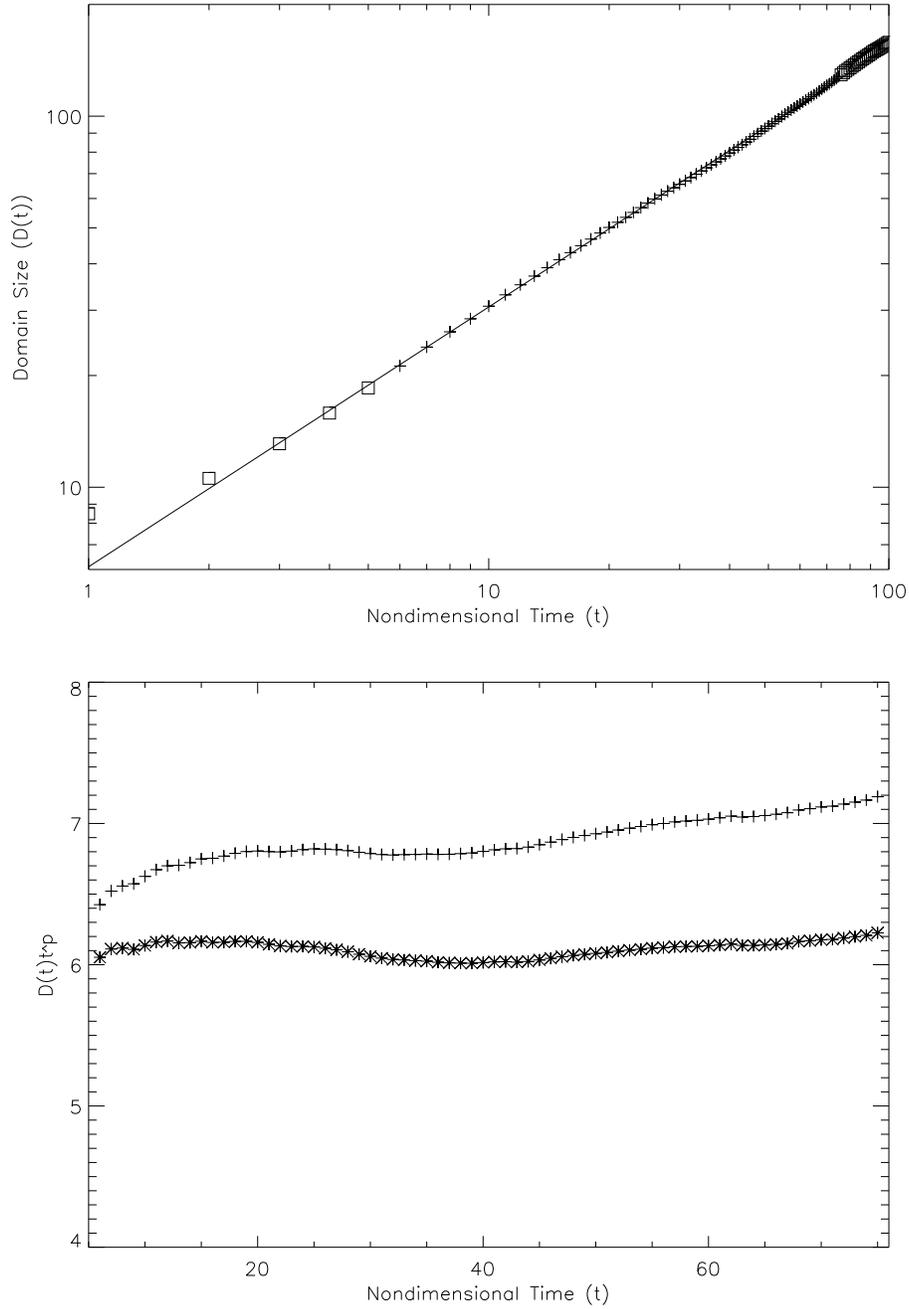

FIG. 6 The domain size as a function of time on a log-log scale. The data is obtained by an ensemble average over 12 different runs, one of which is shown in Fig.5. The runs vary only in the seed used for the random number genrator used to initialize the noise in the density field at time 0. The solid line is the least-squares linear fit obtained for the data with domain sizes lying between $20\Delta x$ and $128\Delta x$. It has a slope of 0.70 with a standard deviation of 0.01. The data used to obtain the fit is shown using '+' and the rest using open squares. For the region of the data fit, in (b), we have plotted $D(t)t^{-2/3}$ in the '+' symbols and $D(t)t^{-0.7}$ in the '*' symbols, where D(t) is the average domain size at time t.

## Acknowledgements

BTN would like to thank Renée Gatignol for making possible his stay at the Laboratoire de Modélisation en Mécanique, Université Pierre et Marie Curie (Paris 6), where part of this work was carried out. Part of this work was carried out under the auspices of the U.S. Department of Energy's CHAMMP program at the Los Alamos National Lab. S. Zaleski would like to thank the programme d'études en microgravité of Centre National d'Etudes Spatiales for its support.

## References

F. J. Alexander, S. Chen, & D.W. Grunau, 1993: Hydrodynamic spinodal decomposition: Growth kinetics and scaling functions. *Phys. Rev. B*, **48**, 634.

C. Appert, J. Olson, D. H. Rothman, & S. Zaleski, 1995: Phase separation in a three-dimensional, two-phase, hydrodynamic lattice gas. *to appear in J. Stat. Phys.*, , .

G.K. Batchelor, 1967: An introduction to fluid dynamics. Cambridge University Press, 60–63.

J.W. Cahn & J.E. Hilliard, 1958: Free energy of a nonuniform system, I. Interfacial energy. *J. Chem. Phys.*, **28**, 258.

J.U. Brackbill, D.B. Kothe, & C. Zemach, 1992: A Continuum Method for Modeling Surface Tension. *J. Comp. Phys.*, **100**, 335–354.

G. Caginalp & E.A. Socolovsky, 1991: Computation of sharp phase boundaries by spreading: The planar and spherically symmetric cases. *J. Comp. Phys.*, **95**, 85–100.

P. Casal & H. Gouin, 1975: Relation entre l'equation de l'énergie et l'équation de mouvement en théorie de Korteweg de la capillarité. *C. R. A. S., ser II*, **300**, 231–234.

J.E. Dunn & J. Serrin, 1965: On the thermodynamics of intersticial working. *Arch. Rational Mech. Anal.*, **88**, 95–133.

R. Evans, 1979: The nature of the liquid–vapour interface and other topics in the statistical mechanics of non-uniform, classical fluids. *Adv. in Phys.*, **28**, 143–200.

F. Falk, 1992: Cahn-Hilliard theory and irreversible thermodynamics. *J. Non-Equilib. Thermodyn.*, **17**, 53–65.

J.E. Farrell & O.T. Valls, 1990: Spinodal decomposition in a two-dimensionall fluid model: Heat, sound, and universality. *Phys. Rev. B*, **42**, 2353–2362.

B.U. Felderhof, 1970: Dynamics of the diffuse gas-liquid interface near the critical point. *Physica*, **48**, 541–560.

H. Furukawa, 1985a: Effect of inertia on droplet growth in a fluid. *Phys. Rev. A*, **31**, 1103–1108.

H. Furukawa, 1985b: Dynamic scaling assumption for phase separation. *Adv. Phys.*, **34**, 703.

J.D. Gunton, M. San Miguel, & P.S. Sahni 1983: in *Phase transitions and critical phenomena*, edited by C. Domb & J.L. Lebowitz, Academic Press, New York.

J.O. Hirschfelder, C.F. Curtiss, & R.B. Bird, 1954: Molecular theory of gases and liquids. John Wiley, New York, chapter 5.

D. D. Joseph, 1990: Fluid mechanics of two miscible liquids with diffusion and gradient stresses. *Euro. J Mech. B: Fluids*, **9**, 565–596.

D.J. Korteweg, 1901: Sur la forme que prennent les équations du mouvement des fluides si l'on tient compte des forces capillaires causées par des variations de densité considérables mais continues et sur la théorie de la capillarité dans l'hypothèse d'une variation continue de la densité. *Archives Néerlandaises des Sciences exactes et naturelles*, **II, 6**, 1–24.

B. Lafaurie *et al.*, 1994: Modelling merging and fragmentation in multiphase flows with SURFER. *J. Comp. Phys.*, **113**, 134-147.

R. Peyret & T.D. Taylor, 1983: Computational methods for fluid flow. *Springer series in computational physics*, Springer-Verlag, 317–319.

T.M. Rogers, K.R. Elder, & R.C. Desai, 1988: Numerical study of the late stages of spinodal

decomposition. *Phys. Rev. B*, **37**, 9638–9649.

P. Seppecher, 1987: Étude d'une modélisation des zones capillaires fluides: interfaces et lignes de contact. Thèse en Mécanique, Université Pierre et Marie Curie (Paris 6).

Michael R. Swift, W. R. Osborn & J. M. Yeomans, 1995: Lattice Boltzmann Simulation of Non-Ideal Fluids. *Phys. Rev. Lett.*, **75**, 803.

J.D. van der Waals, 1894: Thermodynamische Theorie der Kapillarität unter Voraussetzung stetiger Dichteanderung. *Z. Phys. Chem*, **13**, 657.; English translation in *J. Stat. Phys.*, **20**, 197, French translation in Archives Néerlandaises des Sciences exactes et naturelles, **28**,121–209

## Appendix

We recall the form of the mass, momentum and energy conservation equations for a fluid whose free energy is given by the Van der Waals-Cahn-Hilliard model. The mass and momentum equations are identical to those in Section 1. In addition there is the energy equation

$$\partial_t(\rho e_t) + \partial_j(\rho e_t u_j) = \partial_j(\sigma_{ij} u_j) - \partial_j q_j \tag{12}$$

where $\sigma_{ij} = \sigma_{ij}^{(v)} + \sigma_{ij}^{(1)} + p_0 \delta_{ij}$, $e_t = e + u^2/2$ is the total specific energy and $q_j$ is an unknown energy flux. Next, we postulate an extensive (Helmholtz) free energy $F$ in the form

$$F = F_0 + \frac{1}{2}\lambda(\rho, T)|\nabla \rho|^2, \tag{13}$$

where $F_0$ is the bulk free energy. We have generalized the approach of the text to allow for a dependence of $\lambda$ on density and temperature. ¿From thermodynamics, considering $\rho$ and $T$ as the two independent variables of state, the entropy is given by $S = -\partial F/\partial T$, the derivative being taken at constant $\rho$. The extensive internal energy is $\rho e = E$, and is expressed in terms of the free energy by

$$E = F + TS.$$

With these definitions, it may be argued [Dunn and Serrin 1965; Felderhof 1970; Casal and Gouin 1975; Evans, 1979, Falk 1992] that the energy flux and the stresses must be of the form

$$q_i = -k\partial_i T + \rho\lambda\nabla \cdot \mathbf{u}\partial_i\rho, \tag{14}$$

$$\sigma_{ij}^{(1)} = -\lambda \partial_i \rho \partial_j \rho + \frac{1}{2}\left(\lambda - \rho \frac{\partial \lambda}{\partial \rho}\right)|\nabla \rho|^2 + \rho \nabla \cdot (\lambda \nabla \rho)$$

where $k$ is a positive transport coefficient, the $\sigma_{ij}^{(v)}$ are viscous stresses and the bulk pressure $p_0$ is given by $p_0 = -F_0 + \rho \partial F_0/\partial \rho$. The argument is in fact a proof in the perfect fluid, no dissipation case and rests on the usual assumption for entropy production in the dissipative case. We may chose the free energy to be in the van der Waals form

$$F_0 = -\rho R_g T \log\left(\frac{1}{\rho} - b\right) - a\rho^2$$

so that the equation of state is in the familiar form

$$\left(p + \frac{a}{v^2}\right)(v - b) = R_g T. \tag{15}$$

The equation of state and the expression of the stresses are now identical to those in the text if we fix $\theta = T$ and make $\lambda$ a constant, then rewrite equation (15) in the form (5).